# Sifting out communities in large sparse networks

Sharlee Climer[1], Kenneth Smith Jr[1], Wei Yang[2], Lisa de las Fuentes[3,4], Victor G. Dávila-Román[3,5], and C. Charles Gu[4]

[1] Department of Computer Science, University of Missouri – St. Louis, MO, USA

[2] Department of Genetics, Washington University School of Medicine, MO, USA

[3] Cardiovascular Imaging and Clinical Research Core Laboratory, Cardiovascular Division, Department of Medicine, Washington University School of Medicine, MO, USA

[4] Center for Biostatistics and Data Science, Institute for Informatics, Data Science, and Biostatistics (I2BD), Washington University in St. Louis, School of Medicine, St. Louis, MO, USA

[5] Global Health Center, Institute for Public Health, Washington University School of Medicine, MO, USA

Abstract

Research data sets are growing to unprecedented sizes and network modeling is commonly used to extract complex relationships in diverse domains, such as genetic interactions involved in disease, logistics, and social communities. As the number of nodes increases in a network, an increasing sparsity of edges is a practical limitation due to memory restrictions. Moreover, many of these sparse networks exhibit very large numbers of nodes with no adjacent edges, as well as disjoint components of nodes with no edges connecting them. A prevalent aim in network modeling is the identification of clusters, or communities, of nodes that are highly inter-related. Several definitions of 'strong' community structure have been introduced to facilitate this task, each with inherent assumptions and biases. Modularity has gained popularity in recent years due to its ability to identify clusters with non-spherical configurations. With the goal of minimizing partiality, we introduce an intuitive objective function for quantifying the quality of clustering results in large sparse networks. We utilize a two-step method for identifying communities which is especially well-suited for this domain as the first step efficiently divides the network into the disjoint components, while the second step optimizes clustering of the produced components based on the new objective. Using simulated networks, optimization based on the new objective function consistently yields significantly higher accuracy than those based on the modularity function, with the widest gaps appearing for the noisiest networks. Additionally, applications to benchmark problems illustrate the intuitive correctness of our approach. Finally, the practicality of our approach is demonstrated in real-world data in which we identify complex genetic interactions in large-scale networks comprised of tens of thousands of nodes. Due to the resolution limit, modularity fails to partition obvious clusters, while our objective neatly segregates compact clusters for these networks. Based on these three different types of trials, our results clearly demonstrate the usefulness of our two-step procedure and the accuracy of our simple objective.

**Introduction**

Massive datasets are being produced in virtually every field of the sciences, industry, and government. For example, recent advances in high-throughput biotechnology have spawned massive datasets, such as the International Genome Sample Resource, which is comprised of more than 125 million markers for each of 3202 individuals (Byrska-Bishop et al., 2021). A complete network comprised of the markers for this dataset would have more than $7.8 \times 10^{15}$ edges. Indeed, these large-scale networks are highly sparse to yield practical memory requirements and feasible computational running times. Unlike their denser counterparts, these sparse networks tend to have huge numbers of nodes with degree zero as well as naturally separated components. In general, traditional network analyses are not designed for these traits and large-scale sparse networks have posed many challenges for analysts and computational theorists in their efforts to extract useful knowledge from these valuable, and ever growing, datasets.

Currently, one of the greatest challenges appears to be the reliable identification of clusters of nodes representing interacting communities of factors within networks. Networks are constructed to abstractly model complex and diverse relationships such as those that arise for phylogenetic hierarchy of species, literature citations, neural network modeling, and protein-protein interactions, to name a few (e.g. (Blow, 2009; Halperin & Eskin, 2004; Khaledian, Brayton, & Broschat, 2020; M. E. J. Newman, 2001, 2006b; Sia, Jonckheere, & Bogdan, 2019; Watts & Strogatz, 1998; Weighill & Jacobson, 2015; Wolfe, Kohane, & Butte, 2005)). Such networks typically model objects of interest as nodes and place edges between pairs of nodes that exhibit an interesting relationship. Sometimes edges are binary, e.g. two individuals are, or are not, friends on Facebook. On the other hand, sometimes edges have weights, e.g. when a correlation measure is calculated between the two objects. Due to the wide variety of relationships being utilized, many networks do not necessarily obey the triangle inequality and may defy representation in a Euclidean space.

For $n$ nodes, there are $(n^2-n)/2$ possible edges in a network and memory is quickly exhausted for even relatively modest datasets. A common strategy is to select a significance threshold and retain only edges that exceed the threshold. In general, most networks representing large numbers of objects are restricted to substantial sparsity due to this practical consideration.

Computation of pair-wise relationships is commonly feasible, even for large networks, and by leveraging the transitivity assumption, these pair-wise computations can yield arbitrarily high-ordered relationships by identifying communities, or clusters, of highly inter-connected nodes within the network.

A given network can be clustered in a variety of ways, depending on the objective of the clustering algorithm and other selections. Existing clustering algorithms impose various assumptions, some of which can be deleterious for large sparse networks with irregular properties, such as non-Euclidean space or non-spherical community structures.

The focus of this manuscript is on improving the objective function and presenting an efficient two-step procedure for clustering large sparse networks. First, we break the network into components that are completely disconnected from one another. Our approach uses a breadth-first search that efficiently reduces the size of the problem with complete confidence that no true cluster is split. Second, we cluster nodes in each component by optimizing a simple objective function. The new objective is intuitively defined to quantify the "goodness" of clustering while reducing potentially problematic biases. This two-step approach enables us to tackle large networks with high accuracy. Using synthetic data, benchmark instances, and real biological networks, we demonstrate that this technique accurately and effectively identifies communities while avoiding some drawbacks introduced by previous methods.

## Background

We begin by describing some useful parameters in network analysis, followed by a brief review of popular methods for identifying communities in networks. In the following, we define a *singleton* cluster to be one that contains only one node, and a *doubleton* cluster to be one that contains exactly two nodes connected by an edge.

An important parameter of a network is its overall *density*, defined by the ratio of the number of edges in the network to that of a complete network with the same number of nodes in which every pair of nodes is connected by an edge. Therefore, a network with $n$ nodes and $m$ edges has a density of $2m/(n^2 - n)$. In general, density can range from 0 to 1.

The *degree* of a node (the number of edges incident to it) is another important parameter. It is related to the density because the expected minimum degree increases with increasing density of the network. The distribution of the degrees of nodes in many non-random networks exhibit an important property called *Power Law*. Let $p_d$ equal the percentage of the nodes with degree $d$. Many non-random networks found in the real world have degree distributions with tails that follow $p_d \sim d^{-a}$ for some constant $a$ (M. E. Newman, 2003). These networks show a characteristic level of structure because nodes with low degrees are much more prevalent than nodes with high degrees. In contrast, in a random network where the edges are randomly placed, the degrees of the nodes follow a binomial distribution (M. E. Newman, 2003). We will take advantage of the structure of non-random networks in the initial step of our two-step procedure.

Several clustering methods are in wide use in the research community, such as hierarchical clustering, k-means, and modularity. For hierarchical clustering, the implicit objective is to maximize intra-cluster tightness; whereas for *k*-means clustering, the objective is to maximize intra-cluster tightness while maintaining a good distance between clusters (Jain, 2010; Murtagh, 1983). Although not mandated, hierarchical clustering and *k*-means implicitly assume a Euclidean space in which the triangle inequality holds. Furthermore, both methods may have undesirable results for elongated or other irregularly shaped clusters. In other words, the cluster diameters (the maximum distance between any two nodes in the cluster), and/or distances of objects from their respective cluster centers, are to be minimized. This sphericity assumption can lead to incorrect results when the actual clusters are elongated or have irregular shapes (see **Fig S1** of **Supplementary Material** for an example, where water droplets in elongated clouds may be closer to the center of a different cloud than their own).

Newman and Girvan introduced the modularity function $Q$, which avoids these assumptions, and this objective has been rapidly adapted in many domains (M. Newman & Girvan, 2004). Given a symmetric $k \times k$ matrix $e$, where $k$ equals the number of clusters and $e_{ij}$ equals the fraction of edges spanning between clusters $i$ and $j$, modularity is defined as: $Q = \sum_i (e_{ii} - a_i^2)$, where $a_i = \sum_j e_{ij}$. The first term is equal to the fraction of *intra-cluster* edges, which connect nodes within a cluster that is shared by both nodes. The second term represents the fraction of intra-cluster edges that would be expected for random networks with the same number of nodes and corresponding degrees. Identifying communities in a network may be achieved by maximizing the modularity function $Q$, which may require exponential time in the worst case (Brandes et al., 2006).

Several popular implementations using $Q$ exist including a greedy, divisive algorithm (M. Newman & Girvan, 2004), a divisive extremal optimization approach (Duch & Arenas, 2005), and typically slow simulated annealing approaches (Guimerà & Nunes Amaral, 2005). A fast, but less accurate, agglomerative hierarchical clustering algorithm was developed at $O(m\,g \log n)$ computational cost, where $g$ is the depth of the dendrogram and $n$ the number of nodes (Clauset, Newman, & Moore, 2004). Moreover, a number of spectral partitioning methods have been implemented, yielding a balance of time and accuracy (e.g. (M. E. J. Newman, 2006b; Ruan & Zhang, 2008)). The performances of sixteen different implementations based on the optimization of $Q$ have been compared and summarized (Danon, Duch, Diaz-Guilera, & Arenas, 2005).

Fortunato and Barthelemy identified a resolution limit of the modularity $Q$ objective and showed how two clusters with densities of one and connected by a single edge could be merged together into a single cluster when optimizing this function (Fortunato & Barthélemy, 2007). The limit applies to a general class of methods, which is called the *q*-state Potts community detection method and takes modularity as a special case, and depends on the total size of the network (Kumpula, Saramäki, Kaski, & Kertész, 2007). Miyauchi and Kawase proposed a method using modularity to calculated a Z score called Z-modularity. While the resolution limit in the ring of clique problems is overcome by Z-modularity, a network with two pairwise identical cliques exhibits the same resolution limit a modularity (Miyauchi A, Kawase Y 2016).

We have observed that, regardless of the total size of the network, $Q$ has a bias against singleton clusters. Note that the first term of each summand in the $Q$ function is zero when the corresponding node is a singleton cluster, and the second term is always a nonnegative value to be subtracted. Thus, the contribution of a singleton cluster to the $Q$ value is zero if the node is completely isolated, with a degree of zero, and it reduces the value of $Q$ in all other cases. As a result, $Q$ consistently favors retaining nodes with low degrees within a larger community rather than cutting them away into singleton clusters. **Figure 1** shows a toy example demonstrating this behavior. We have also observed a potential bias against doubleton clusters while using $Q$. This is demonstrated in **Figure S2** of the **Supplementary Material**.

In many real-world situations modeled as sparse networks, we expect large numbers of singleton and doubleton clusters. Due to the Power Law, large sparse networks are expected to produce a sizeable percentage of nodes with a degree of zero. Most network models include a fair number of noisy edges and these edges may attach zero-degree nodes to clusters or present as doubletons in the network. On the other hand, doubletons may also naturally arise due to a strong pair-wise relationship, such as two genes that express together to produce a protein. In short, bias against singleton or doubleton clusters hold potential to encumber research advancements utilizing large sparse networks.

## Methods and Materials

We present a two-step procedure, called *Sieve*, that achieves: (1) efficient separation of disjoint components in large sparse networks, and (2) optimal identification of potential functional units by further "cutting" of the connected components into communities using a simple objective function, as described in the following.

***Step 1: Search for isolated components***. In large sparse networks it is advantageous to first separate out isolated components that are disconnected from each other in the network. We implemented a procedure based on breadth-first search (BFS) to identify such isolated network components.

The BFS starts at an arbitrary node and finds every other node that is reachable from this node. These nodes comprise one connected component. Then BFS starts at a node that hasn't been visited yet and repeats the search. This process is repeated until all nodes have been visited. At this point, all of the components are identified, and there are *no* edges spanning between the identified components. Hence, for the given network, we have complete confidence that true clusters were not split and BFS essentially divides-and-conquers large sparse networks without any loss of accuracy. Our implementation of BFS uses linked lists to take advantage of sparsity and doesn't store the search tree, requiring only $O(n + m)$ of space, where $n$ is the number of nodes and $m$ is the number of edges. The time complexity is also $O(n + m)$.

***Step 2: Extraction of communities in the identified components using a novel objective function***. We have developed a quality measure, $S$, that is suitable for sparse networks as follows:

$$\max S = \sum_{i=1}^{C} {n_i}/{n} \, S_i \qquad \text{(eq. 1)}$$

$$\text{where } S_i = \sum_{j=1}^{k_i}(o_{ij} - r_{ij}) \qquad \text{(eq. 2)}$$

$$\text{and } k_i = 1, \text{ for } d_i \geq D \qquad \text{(eq. 3)}$$

where $C$ = the number of connected components, $n_i$ = the number of nodes in component $i$, $n$ = the number of nodes in the entire network, $k_i$ = the number of clusters in component $i$, $o_{ij}$ = fraction of observed edges that are intra-cluster edges in cluster $ij$, $r_{ij}$ = the expected fraction of edges completely within a cluster with the same number of nodes as cluster $ij$ for a random component with the same density, $d_i$ = the density of component $i$, and $D$ is a user-adjustable parameter, with a default value of 0.5 (this default is used for all of our trials presented in this manuscript).

In essence, $S_i$ simply quantifies the difference between the number of intra-cluster edges present and the number expected in clusters containing the same numbers of nodes in a random component with an equal density. The overall $S$ value is the weighted average of the $S_i$ values. Given a network with $n > 1$ nodes, $m > 0$ edges, and a partitioning into $C$ connected components, the following properties hold:

***Theorem 1***. $-1 < S_i < 1$ for $1 \leq i \leq C$.

***Theorem 2***. The maximum $S_i$ value for component $i$ is at least zero for $1 \leq i \leq C$.

These theorems are proved in the **Supplementary Material**. Note that when $k_i$ is equal to $n_i$, each node is a singleton cluster and $S_i = 0$. At the other extreme, when $k_i$ is equal to 1 and all nodes are contained in a single cluster, $S_i$ is also equal to zero. $S_i > 0$ indicate more informative clustering arrangements of the component and $S_i < 0$ arises when the clusters contain fewer edges than expected by random chance.

Importantly, we minimize the risks of reaching the resolution limit by optimizing each component using the local number of nodes, rather than the global number of nodes in the entire network, $n$. As $n$ grows for large networks, modularity $Q$ is biased against the removal of edges, even if a single edge connects two dense clusters. $S$ utilizes the local number of nodes in a given component in the $r_{ij}$ term, thereby alleviating this issue.

We now cast the problem of maximizing $S$ as a mixed-integer linear program (MIP). Let $m_i$ = the number of edges in component $i$, $m_{ij}$ equal the number of intra-cluster edges in cluster $ij$, and $n_{ij}$ equal the number of nodes in cluster $ij$. It follows that $o_{ij} = m_{ij}/m_i$ and $r_{ij} = (n_{ij}^2 - n_{ij})/(n_i^2 - n_i)$. Let $b_i$ equal the fraction of edges that span between clusters (inter-cluster edges) for component $i$. Since the fraction of intra-cluster edges plus the fraction of inter-cluster edges is equal to one, $\sum_{j=1}^{k_i} o_{ij} + b_i = 1$, it follows that $\sum_{j=1}^{k_i}(o_{ij} - r_{ij}) = 1 - \sum_{j=1}^{k_i} r_{ij} - b_i$. While this formulation of $S_i$ is less intuitive, it yields efficient maximization. In addition, the quadratic terms in $r_{ij}$ can be linearized using the fact that $n_i$ is a constant and that $n_{ij}^2 - n_{ij}$ can be calculated by linear terms as $n_{ij}^2 - n_{ij} = n_{ij}(n_{ij} - 1) = 2\sum_{a=1}^{n_{ij}-1} a$.

The algorithms are implemented in a C++ program called *Sieve*. We evaluate the *Sieve* method by performance comparison with modularity. Due to the fact that the components identified by BFS are independent of each other, the $Q$ and $S$ objective functions can be solved independently for each component, with the results combined for the overall optimal solution. We have cast both $Q$ and $S$ as MIPs and solved to optimality using IBM's Cplex. Derivations of the formulae and the pseudocode for the algorithms are supplied in **Supplementary Material**. Optimally solving the MIPs may require exponential time in the worst case, but we have successfully computed globally optimal solutions for all of our experiments using Cplex version 12.10 (https://www.ibm.com/analytics/cplex-optimizer). Application of $S$ for the toy example is shown in **Figure 1(d)** where all clusters are correctly identified except one singleton.

We use simulated networks with known clusters to evaluate the accuracies of the methods. We then compare the performances of both methods for benchmark instances and two biological networks. In all trials, breadth-first search (BFS) in Step 1 is applied first to identify disjoint components. For each component with density < 0.5, clusters within the component are identified by optimally solving the corresponding MIP for $S$. All evaluation analyses were carried out using Ubuntu 20.04.1 LTS 64-bit on Intel Core i5-4690K @ 3.50 GHz x4, with 16 GB memory.

We attempted generating the synthetic networks using the popular LFR generator (Lancichinetti, Fortunato and Radicchi, 2008). However, the LFR generator could not produce networks that met the characterists observed in the biological data sets, primarily the large number of singletons and doubletons. Consequently, we simulated networks by adding random edges to ground-truth clusters with properties similar to the biological networks of interest. Our approach involves three parameters with clear impacts on network structure and we generated datasets by using every possible combination of settings for the parameters. We first generated networks with $n$ = 10,000 nodes and 200 non-singleton ground-truth clusters. Since we have observed a large percentage of doubleton clusters in biological data, half of these clusters are doubletons. The other half of the clusters have a random number of nodes, ranging from 3 to 10; and of densities randomly chosen between 0.8 and 1.0. The resulting ground-truth clusters are stored then noisy edges are added as follows. We add "bridges" between non-singleton clusters randomly selected at a rate of $\beta$ x $k$, where $\beta$ = 0.05 and 0.10, and $k$ = 200 (the number of non-singleton clusters). The number of edges in these bridges is equal to the number of nodes in the smaller cluster being bridged times $\varepsilon$, which is set to either 0.2 or 0.5. Finally, we add completely random edges to the network such that the percentage of these edges $r$ = 0%, 5%, and 10% of the total number of edges in the network, $m$. Every combination of these parameters is applied, yielding 12 scenarios. 100 random instances were generated for each of the 12 scenarios, yielding 1200 test datasets with known ground truth.

The accuracies of the methods were measured by agreement between communities detected by the methods and the original ground-truth communities using the Jaccard Index (JI) (Jaccard, 1901). If no pairs of nodes are grouped into the same cluster in both the true and computed results, JI = 0. If the clustering results are identical, JI = 1. For example, the JI values for the toy problem solutions shown in **Figure 1 (c)** and **(d)** are 0.491 (for $Q$) and 0.961 (for $S$). The JI algorithm is given in the **Supplementary Material**.

We also compare $Q$ and $S$ for seven popular benchmark instances. Networks tested include karate club (Zachary, 1977), dolphins (Lusseau et al., 2003), college football (Girvan & Newman, 2002), Les Miserables (Knuth, 1994), David Copperfield word adjacencies (M. E. J. Newman, 2006a), Krebs' political books (http://www.orgnet.com/), and Chesapeake Bay ecosystem (Baird & Ulanowicz, 1989). Additional descriptions of these instances are given on the DIMACS Challenge report (Bader et al., 2017) and in the **Supplementary Material**.

Finally, we compare $Q$ and $S$ for two biological networks. Levels of gene expression for individuals exhibiting Alzheimer's disease (AD) and normal controls, as well as gene expression for individuals with and without various types of influenza were used to create the two networks. These data were downloaded from NIH's Gene Expression Omnibus (https://www.ncbi.nlm.nih.gov/gds), accession numbers GSE15222 and GSE68310, respectively. The AD network is comprised of 17,120 nodes with correlations computed across 364 individuals, while the influenza network is comprised of 94,208 nodes and based on 880 individuals. For each network, correlations were computed for every pair of genes and the highest 1000 correlations were represented by edges in the network. Cleaned data and network files are available from the corresponding authors by request.

## Results

In this section, we present results for synthetic networks, benchmarks, and biological data. Using the 12 scenarios shown in **Fig. 2**, we generated 100 random synthetic networks per scenario and optimally solved $Q$ and $S$ for each network, as described in the last section. As shown in **Fig. 2**, $S$ was significantly more accurate than $Q$ for all scenarios. The gap is widest for scenarios 9 and 12, which include 10% random edges.

We then computed optimal solutions for both $Q$ and $S$ for seven popular benchmark instances. The results for Zachary's karate club are shown in **Fig. 3** and the plots for the other instances are supplied in the **Supplementary Material**. Optimizing $S$ yielded two singleton clusters, as shown in **Fig. 3**. The only other difference between the $Q$ and $S$ results is that one node is assigned to different clusters. The $S$ clustering results with one more intra-cluster edge and one less inter-cluster edge than the $Q$ result. Increasing intra-cluster edges and decreasing inter-cluster edges is a common goal for clustering in general.

Finally, we computed optimal solutions for the two biological networks described in the previous section. The size distributions of the components identified using BFS is shown in **Fig. 4**. Both networks are sparse, with more than 97% of the nodes having degrees of zero. The Alzheimer's network was comprised of 31 components with only one edge (doubletons) and 24 components with more than one edge. The influenza network included 79 doubletons and 36 with more than one edge.

One component from the influenza network is shown in **Fig. 5** and additional components are included in the **Supplementary Material**. Overall, $Q$ only broke up five components into clusters, three in the influenza network and two in the Alzheimer's network, likely due to the resolution limit. On the other hand, $S$ split 25 components, 14 and 11 in influenza and Alzheimer's networks, respectively, presumably due to its employment of the local number of nodes in its formulation.

Running BFS required 0.15 seconds on the largest network tested (influenza). Interestingly, our Cplex implementations ran more quickly for $Q$ than for $S$. For example, the karate club instance with 34 nodes and 78 edges required only 0.09 seconds to find the optimal solution for $Q$ and almost two minutes for optimally computing $S$.

## Discussion

We have introduced a two-step procedure, called *Sieve*, for identifying clusters in sparse networks containing disjoint components. First, we accurately identify these disconnected components using breadth-first search (BFS), then we use a novel objective function, $S$, to search for clusters within the components. The BFS implementation divides extremely large but very sparse networks into less computationally-intensive disconnected components, and the simple $S$ objective is not affected by the shapes of clusters, is appropriate for Euclidean or non-Euclidean spaces without regard for obeyance of the triangle inequality, is not biased against singleton and doubleton clusters, and alleviates the resolution limit by utilizing the local, rather than global, number of nodes in the network.

Using synthetic networks, for which the true clusters are known, $S$ exhibited substantially higher accuracy than $Q$ over 100 trials for each of twelve scenarios. Computations using benchmark instances presented intuitive evidence supporting the higher accuracy of $S$ by making a node assignment for Zachary's karate club that results with one more intra-cluster edge and one less inter-cluster edge than the assignments for the $Q$ optimization. Finally, trials on two large real-world biological networks confirmed that the resolution limit suffered by $Q$ is not exhibited for $S$.

Other aspects of the *Sieve* method may be further developed to improve its performance. In our trials, the largest component contained 163 nodes. For substantially larger component sizes, it may not be feasible to compute the optimal solution. In these cases, the 'anytime' solution provided by Cplex when the computations are terminated can be used as an approximate solution. A bonus is that Cplex is able to provide bounds on the quality of these anytime solutions. However, excessively large components will exhaust memory fairly early in the search. In addition to developing strategies to speed up our Cplex implementation, we are currently developing an approximate solver using a genetic algorithm. This program will be able to scale to large instances within reasonable time frames, while maintaining limited memory requirements.

We are also working on extending our objective function for weighted networks. In some networks, edges have weights reflecting relationships between pairs of nodes; and some unweighted methods have been extended to weighted networks (e.g. (Ruan & Zhang, 2006)) to leverage this additional information.

Finally, it should be noted that in addition to selection of an appropriate objective function, the choice of relationship measure can have a dramatic effect on network topology, and different thresholds used for these pair-wise relationships result in varying network sparsity levels. Both may drastically affect clustering results, and these choices are generally domain and instance specific.

The *Sieve* method has important implications in analyzing real world data, which are rapidly growing for many critical research areas. For example, most complex diseases of interest arise due to system failures caused by concerted action of many genetic variants, coupled with environmental factors. The cardinality of involved variants may be high, resulting with small marginal effects per associated variant. Moreover, these factors may interact synergistically, for which no marginal effects may be exhibited when examined in isolation. Computing associations for high-ordered interactions is computationally infeasible beyond pairwise in general, and if possible, multiple-testing corrections for such trials would be prohibitive. Consequently, network modeling is an effective and practical approach for this and similar data, such as those that arise in virus transmission predictions, meteorology, subatomic modeling, and environmental monitoring, to name a few. However, identifying and analyzing community relationships hidden in such voluminous numbers of variables is largely an open area. Insidious barriers to research advancements include subtle assumptions that may unfavorably bias results, such as those described herein. This manuscript provides an alternative strategy that avoids these subtle assumptions while sifting communities out of large sparse networks, which are becoming increasingly prevalent in the pursuit of knowledge to improve the health and well-being of the global society.

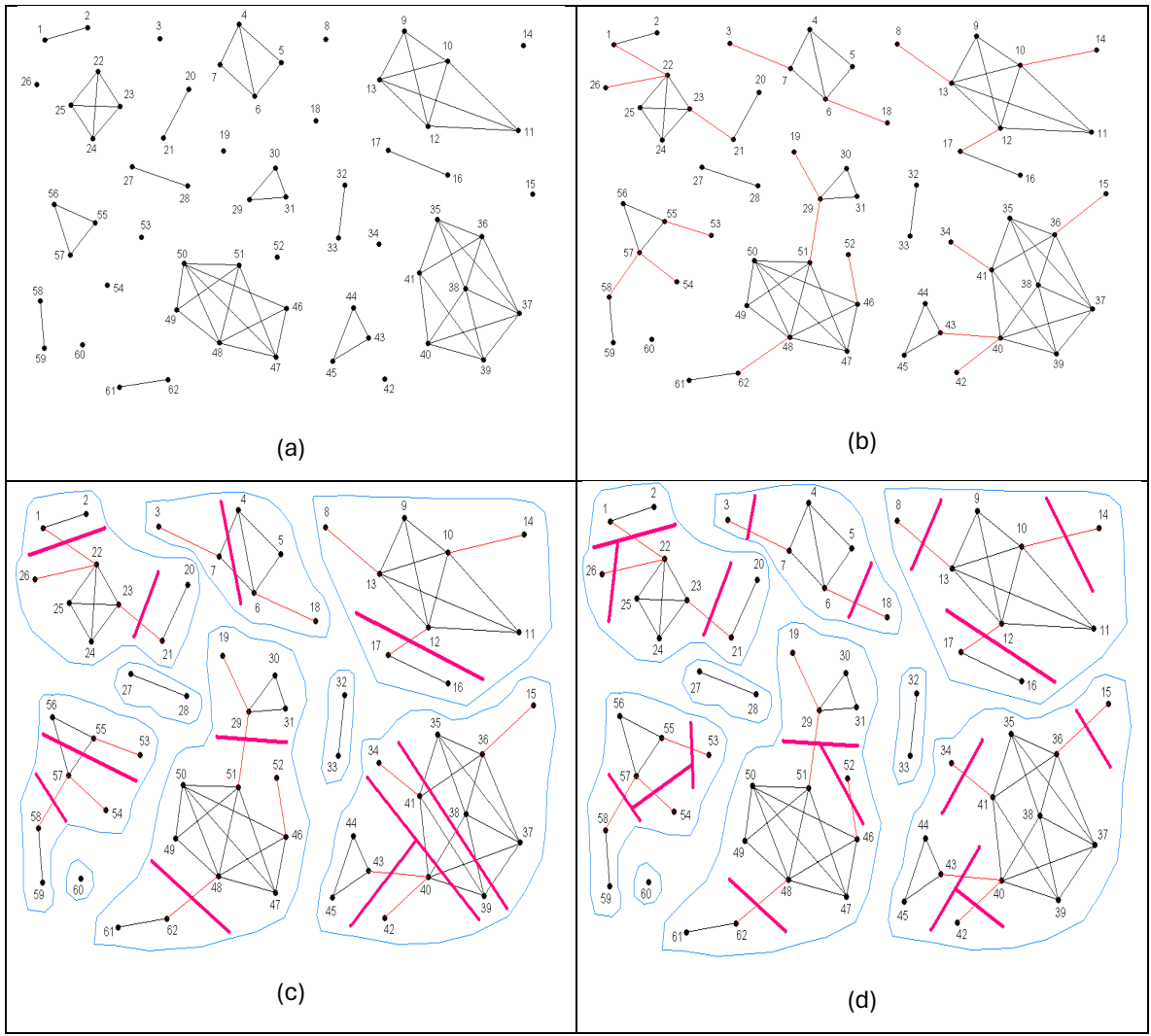

**Figure 1.** (a) A set of clusters. (b) Clusters with noisy (red) edges added. (c) Clustering results when optimizing modularity, $Q$. The light blue lines outline each connected component and the dark red lines indicate cluster memberships. Note that dense clusters are split in order to avoid singleton clusters, especially in the component in the lower right corner. (d) Clustering results when optimizing $S$. All but one singleton node (#19) are identified.

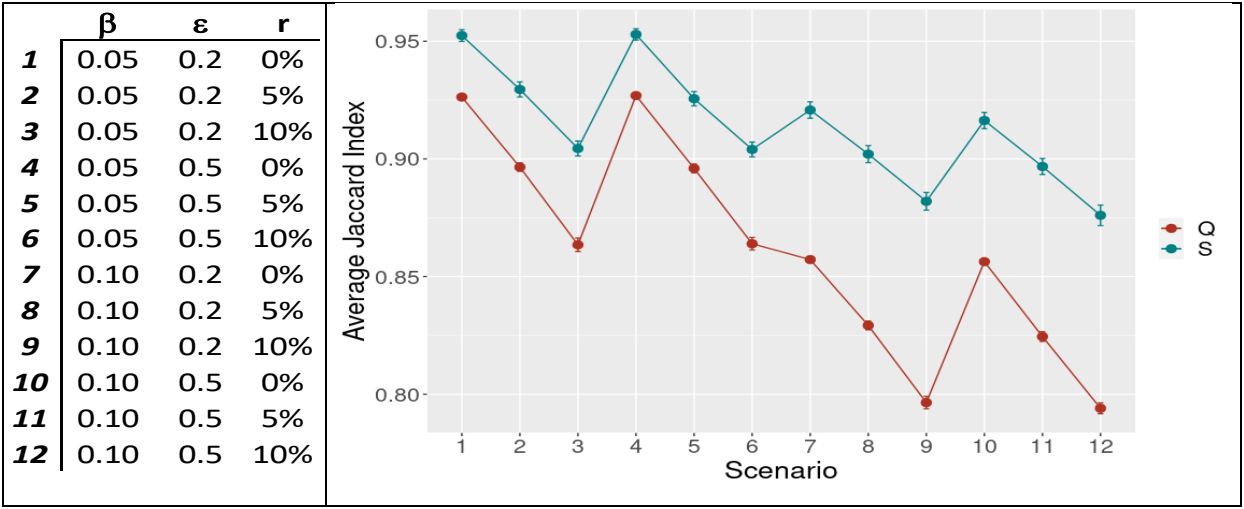

**Figure 2.** Average Jaccard Index for $Q$ and $S$ for the 12 synthetic network scenarios. Bars indicate the 95% confidence interval.

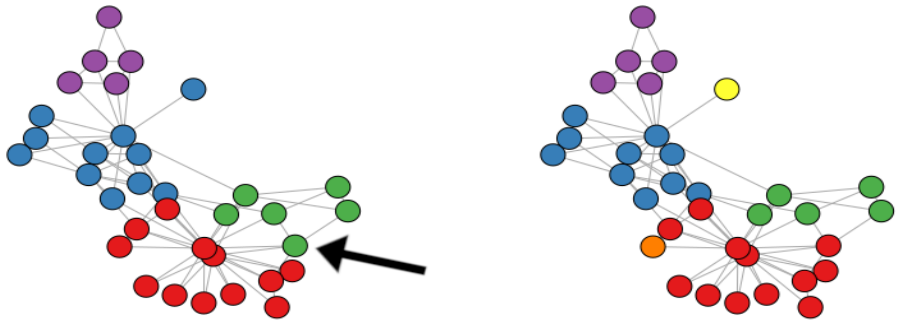

**Figure 3**. Clustering results for Zachary's karate club network for $Q$ on the left and $S$ on the right. Results are identical except for three nodes. Two nodes, each attached by a single edge, were made into singletons by $S$ (yellow and orange nodes on right). Interestingly, $Q$ assigned the node pointed to by the arrow to the green cluster despite the fact that it only has two edges connecting it to that cluster, whereas $S$ assigned this node to the red cluster, which yields three intra-cluster edges.

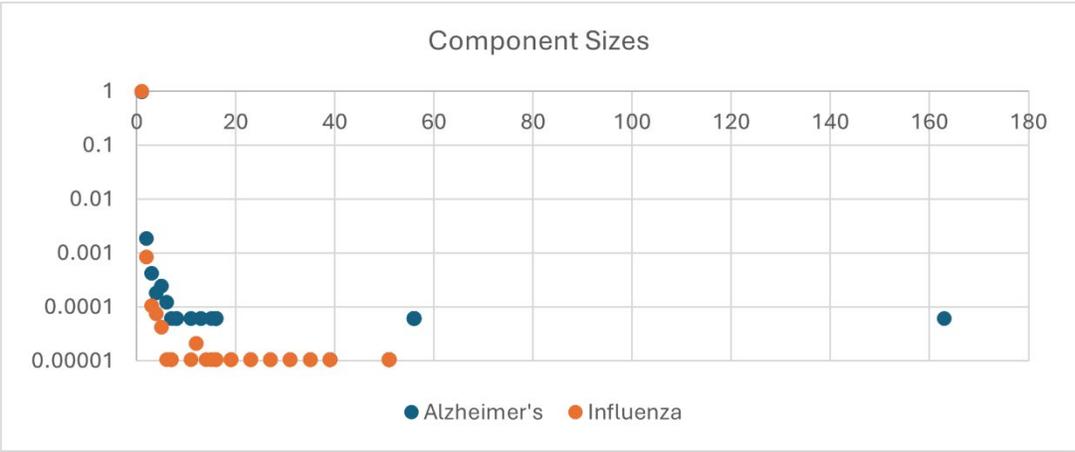

**Figure 4**. Fractions of components with various numbers of nodes (1 to 163) for the Alzheimer's and Influenza networks. 97.5% and 99.4% of the Alzheimer's and Influenza nodes, respectively, had degree of zero.

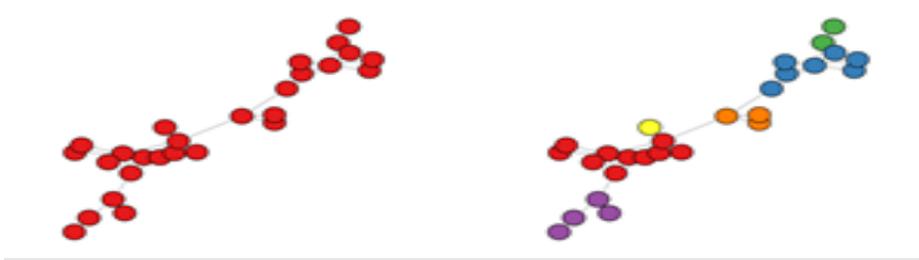

**Figure 5**. Clustering results for the 27-node component from the influenza network. $Q$, on the left, retained all 27 nodes in a single cluster, while $S$, on the right, split it into compact clusters.

# Sifting out communities in large sparse networks

Sharlee Climer[1], Kenneth Smith Jr[1], Wei Yang[2], Lisa de las Fuentes[3,4], Victor G. Dávila-Román[3,5], and C. Charles Gu[4]

[1] Department of Computer Science, University of Missouri – St. Louis, MO, USA

[2] Department of Genetics, Washington University School of Medicine, MO, USA

[3] Cardiovascular Imaging and Clinical Research Core Laboratory, Cardiovascular Division, Department of Medicine, Washington University School of Medicine, MO, USA

[4] Center for Biostatistics and Data Science, Institute for Informatics, Data Science, and Biostatistics (I2BD), Washington University in St. Louis, School of Medicine, St. Louis, MO, USA

[5] Global Health Center, Institute for Public Health, Washington University School of Medicine, MO, USA

# Supplementary Material

## Proofs

Given a network with $n > 1$ nodes, $m > 0$ edges, and a partitioning into $C$ connected components, where $n_i \geq 1$ is the number of nodes in component $i$, the following properties hold:

***Theorem 1.*** $S_i < 1$ for $1 \leq i \leq C$.

***Proof.*** First we prove $S_i < 1$.

$$S_i = \sum_{j=1}^{k_i}(o_{ij} - r_{ij}) = \sum_{j=1}^{k_i}\frac{m_{ij}}{m_i} - \sum_{j=1}^{k_i}\frac{n_{ij}^2 - n_{ij}}{n_i^2 - n_i} \qquad (eq.\,1)$$

where $k_i$ = the number of clusters in component $i$, $m_{ij} \geq 0$ is the number of edges in cluster $j$ of component $i$, and $n_{ij} > 0$ is the number of nodes in cluster $j$ of component $i$. By definition:

$$0 \leq \frac{m_{ij}}{m_i} \leq 1, \text{ for } 1 \leq j \leq k_i \qquad (eq.\,2)$$

$$0 \leq \frac{n_{ij}^2 - n_{ij}}{n_i^2 - n_i} \leq 1, \text{ for } 1 \leq j \leq k_i \qquad (eq.\,3)$$

It follows that $S_i = 1$ if and only if $\sum_{j=1}^{k_i} m_{ij}/m_i = 1$ and $\left.(n_{ij}^2 - n_{ij})\middle/(n_i^2 - n_i)\right. = 0$, and it can never exceed 1. Since $n_{ij} > 0$, the only case in which $n_{ij}^2 - n_{ij} = 0$ is when $n_{ij} = 1$ (a singleton cluster) and is strictly positive for $n_{ij} > 1$. Consequently, the subtrahend of eq. 1 is zero if and only if all of the clusters are singleton clusters. Note that the minuend of eq. 1 is zero for singleton clusters. Therefore, $S_i < 1$.

***Theorem 2.*** The maximum $S_i$ value for component $i$ is at least zero for $1 \leq i \leq C$.

***Proof.*** We show that $S_i = 0$ when $k_i = 1$ to prove there always exists at least one clustering of the nodes that yields a value of at least zero. For $k_i = 1$:

$$S_i = \frac{m_i}{m_i} - \frac{n_i^2 - n_i}{n_i^2 - n_i} = 0 \qquad (eq.\,4)$$

# Mixed Integer Linear Program Formulation

In this section, we formulate the problem of maximizing $S_i$ for component $i$. For simplicity, we drop the $i$ indices. Let $n_j$ equal the number of nodes in cluster $j$ and $n$ equal the number of nodes in component $i$.

Maximizing $S = 1 - \sum_{j=1}^{k} r_j - b$ is equivalent to minimizing $Z = \sum_{j=1}^{k} r_j + b$. Note that $r_j = (n_j^2 - n_j)/(n^2 - n)$ and $b = \frac{e}{m}$, where $e$ is the number of inter-cluster edges and $m$ is the total number of edges in the component. Let $R = \frac{n^2 - n}{m}$. Then minimizing $Z$ is equivalent to minimizing $Z' = \sum_{j=1}^{k} n_j(n_j - 1) + Re$. Note that $n_j(n_j - 1) = 2 \sum_{a=1}^{n_j - 1} a$.

We use the following variables for our formulation. Let $n$ equal the number of nodes in component $i$ and $K$ equal the maximum number of non-singleton clusters in component $i$, $K = \lfloor \frac{n}{2} \rfloor$ to preserve optimality; and $S$ equal the maximum cluster size, $S = n$ to preserve optimality. Let $x_{jk} = 1$ if node $j$ is in cluster $k$, and equal 0 otherwise. Let $c_e = 1$ if edge $e$ is inter-cluster and 0 otherwise. Finally, let $b_{sk}$ be a binary variable representing the $s^{th}$ member of cluster $k$. The clusters are numbered from 1 to $K$. We define cluster $K+1$ as the "cluster" containing only nodes that comprise singleton clusters. Note that some of the other clusters may be singletons also. The integer linear program formulation follows:

$$\min Z' = 2 \sum_{k=1}^{K} \sum_{s=1}^{S} (i-1) b_{sk} + R \sum_{e=1}^{m} c_e$$

such that:

$$\sum_{k=1}^{K+1} x_{jk} = 1 \qquad \text{for } 1 \leq j \leq n \qquad (1)$$
$$n_k = \sum_{j=1}^{n} x_{jk} \qquad \text{for } 1 \leq k \leq K \qquad (2)$$
$$\sum_{s=1}^{S} b_{sk} = n_k \qquad \text{for } 1 \leq k \leq K \qquad (3)$$
$$c_e \geq x_{uk} - x_{vk} \qquad \text{for } 1 \leq k \leq K \quad \forall e = (u,v) \in \text{edges} \qquad (4)$$
$$c_e \geq \frac{1}{2}(x_{u(K+1)} + x_{v(K+1)}) \qquad \forall e = (u,v) \in \text{edges} \qquad (5)$$

Constraints (1) ensure that each node is assigned to exactly one cluster. Constraints (2) require that $n_k$ is the size of cluster $k$. Constraints (3) set the sum of the number of "members" equal to $n_k$. Constraints (4) force $c_e$ to be equal to one if edge $e$ is inter-cluster. Finally, constraints (5) require that $c_e = 1$ if nodes $u$ or $v$ is in "cluster" $K+1$.

# Pseudocode

Breadth-First Search in Step 1:

```
initialize an array visited[n] to all 0's

for each node v
      if (visited[v] = 0)
            put v in a queue
            visited[v] = 1
            while (queue is not empty)
                  -remove node w from queue
                  -add w to component
                  -add all unvisited nodes that are incident to w to
                  queue and mark as visited
            record component
```

Clustering by *S* in Step 2:

```
apply breadth-first search to identify components
number = 1

for each component c
      if (density(c) > threshold)
            for each member i of c
                  cluster[i] = number
            number = number + 1
      else
            buildModel(c)
            solve(c)
            for each cluster k that is identified
                  for each member i of k
                        cluster[i] = number
                  number = number + 1
```

## Jaccard Index Definition

Let $S_t = \{(u, v): u, v \in c_t\}$, where $c_t$ is a true cluster, be the set of all pairs of nodes that appear together in any original cluster. Let $S_c$ be similarly defined for the computed clustering result. Let $N_i$ equal the size of $S_t \cap S_c$, and $N_u$ the size of $S_t \cup S_c$, then the Jaccard Index is calculated by $JI = N_i/N_u$.

## Benchmark Descriptions

| Network | Nodes | Edges | Avg Degree |
|---|---|---|---|
| karate | 34 | 78 | 2.294 |
| chesapeake | 39 | 170 | 4.359 |
| dolphins | 62 | 159 | 2.565 |
| lesmis | 77 | 254 | 3.299 |
| polbooks | 105 | 441 | 4.200 |
| adjnoun | 113 | 425 | 3.795 |
| football | 115 | 613 | 5.330 |

| Network | Nodes | Edges | Q | | S | | |
|---|---|---|---|---|---|---|---|
| | | | Objective | # Clusters | Objective | # Clusters | # Singletons |
| karate | 34 | 78 | 0.419790 | 4 | 0.484437 | 4 | 2 |
| chesapeake | 39 | 170 | 0.265796 | 3 | 0.339851 | 4 | 3 |
| dolphins | 62 | 159 | 0.528519 | 5 | 0.578280 | 6 | 10 |
| lesmis | 77 | 254 | 0.566688 | 6 | 0.644585 | 9 | 8 |
| polbooks* | 105 | 441 | 0.527237 | 5 | 0.602041 | 8 | 0 |
| adjnoun* | 112 | 425 | 0.313367 | 7 | 0.442449 | 10 | 14 |
| football* | 115 | 613 | 0.604570 | 10 | 0.611332 | 10 | 0 |

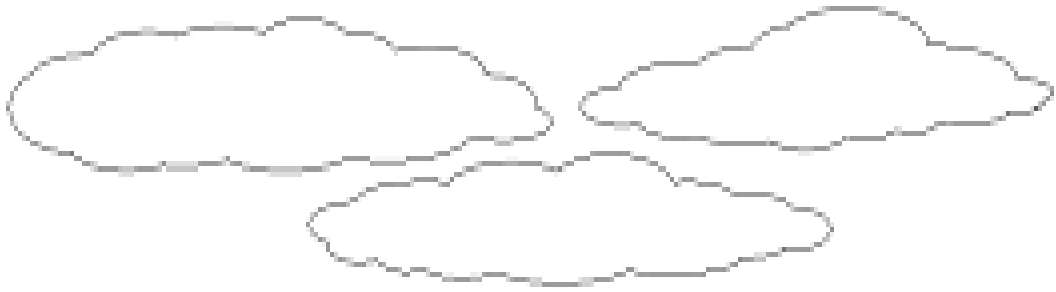

**Figure S1**. Clustering the droplets in the clouds using Euclidean distance and an algorithm that assumed sphericity would give erroneous results for these three clusters as many droplets are closer to the centers of other clouds than the center of their own cloud.

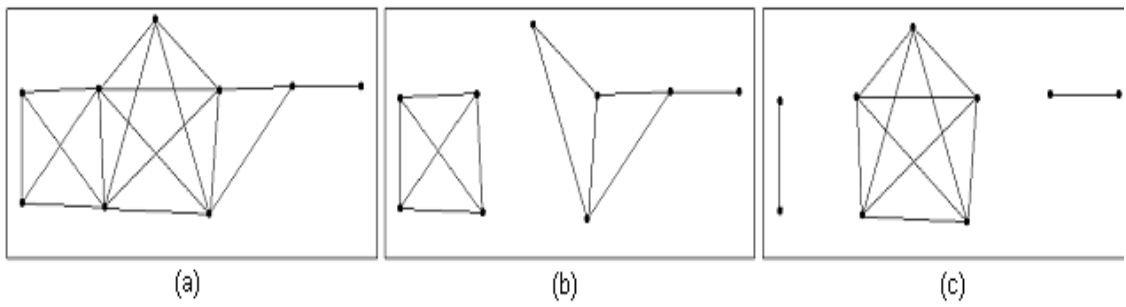

**Figure S2**. (a) A small network. (b) Optimizing Q cuts 6 edges, yielding 2 clusters with densities of 1 and 0.6, respectively. The corresponding Q value is 0.1667. (c) Cutting 6 edges can also yield 3 clusters all with densities of 1. The corresponding Q value is 0.1049. Note that the Q value is dramatically higher for (b), in which doubleton clusters have been eliminated, despite the lower densities exhibited.

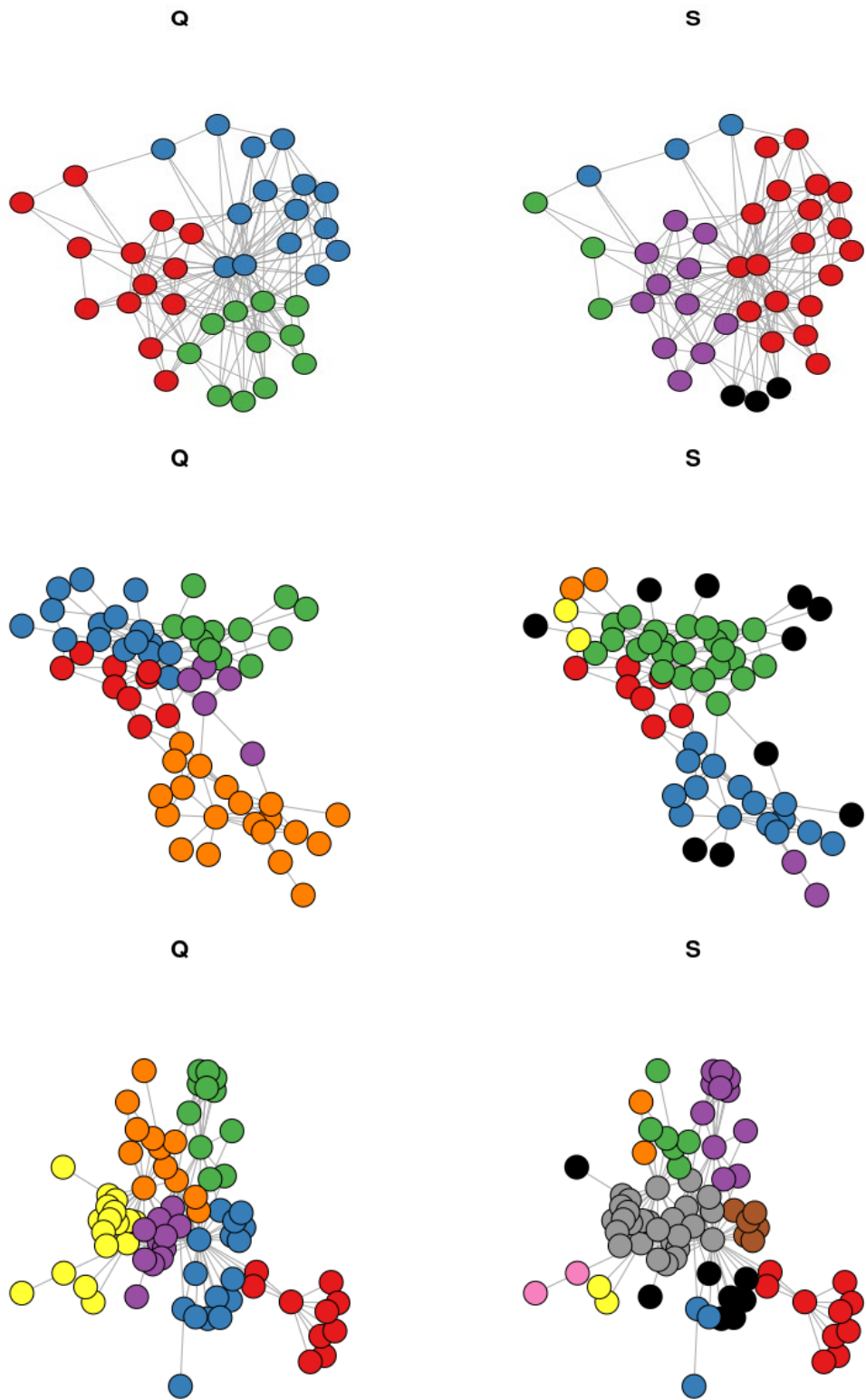

**Figure S3 (part 1).** Plots for (top to bottom) chesapeake, dolphins, and lesmis benchmark instances. Black dots indicate singleton clusters.

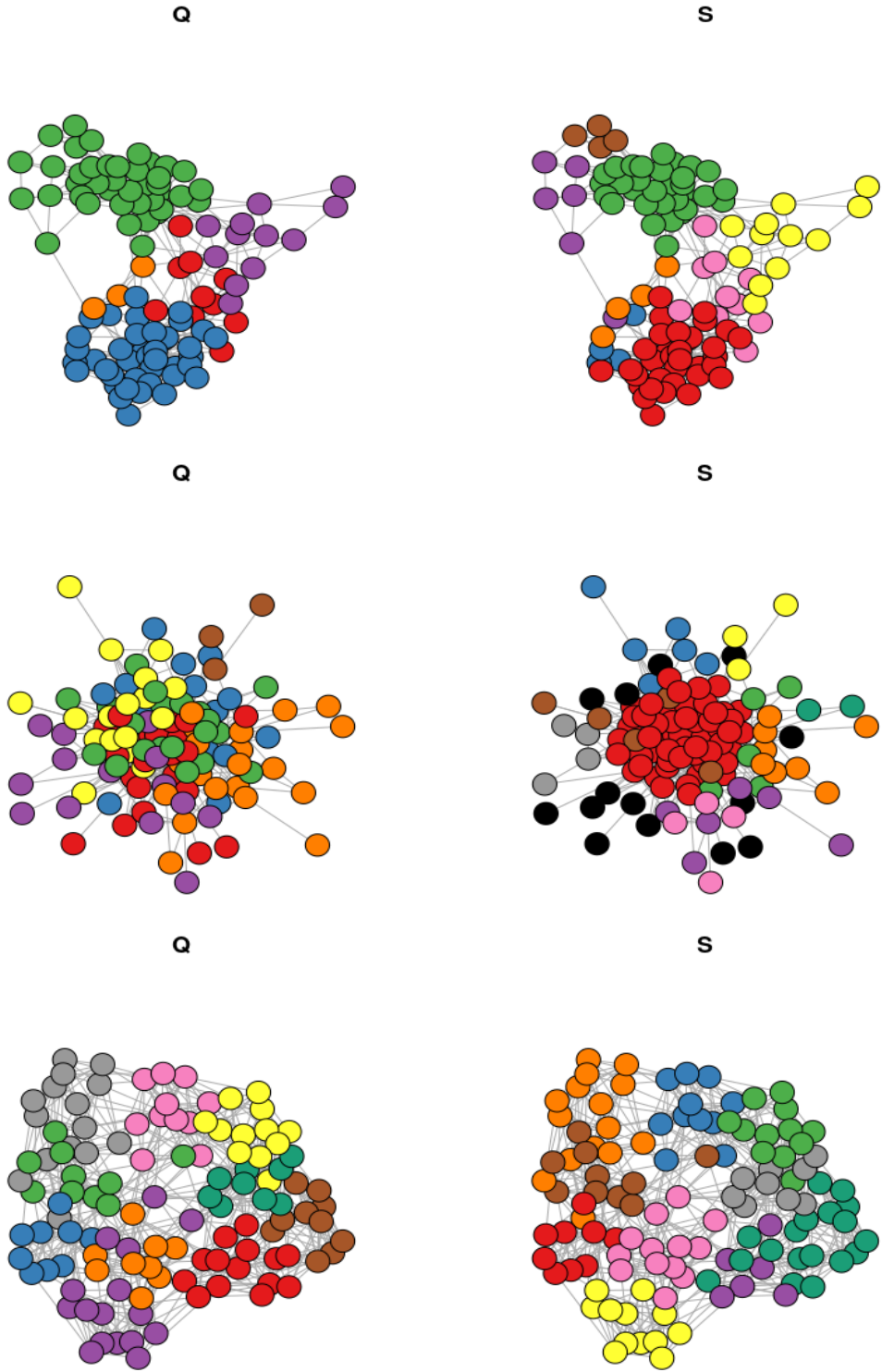

**Figure S4 (part 2)**. Plots for (top to bottom) polbooks, adjnoun, and football benchmark instances. Black dots indicate singleton clusters. Note these plots show anytime solutions for *S*.

10